\documentclass{iaus}
\usepackage{graphicx}

\title[Mergers of luminous early-type galaxies] 
{Mergers of luminous early-type galaxies}

\author[Z. L. Wen, J. L. Han \& F. S. Liu]   
{Z. L. Wen$^1$, J. L. Han$^1$ \and F. S. Liu$^2$}

\affiliation{$^1$National Astronomical Observatories, Chinese Academy of Sciences, China\\
[\affilskip]$^2$College of Physics Science and Technology, 
                 Shenyang Normal University, China} 

\pubyear{2009}
\volume{267}  
\pagerange{119--126}
\jname{Co-Evolution of Central Black Holes and Galaxies}
\editors{B.M.\ Peterson, R.S.\ Somerville, \& T.\ Storchi-Bergmann, eds.}
\begin{document}

\maketitle


\begin{table}[b]
  \begin{center}
  \caption{Close pairs for merger statistics.}
  \label{tab1}
 {\scriptsize
  \begin{tabular}{lrcrrl}\hline
Authors                                & Pairs No.       &Image-check&  Redshift    &   Magnitude range    &  merger type\\
\hline
Carlberg et al. 2000, ApJ 532, L1  &      109 	 & No        &	0.1--1.1       &      $ M_B<-20.5$	 &  general    \\
Patton et al. 2000, ApJ 536, 153   &       80 	 & No        &	0.005--0.05    & $-21.7<M_B<-18.7$	 &  general    \\
Patton et al. 2002, ApJ 565, 208   &       88	 & No        &	0.12--0.55     & $-21.7<M_B<-18.7$	 &  general    \\
Lin et al. 2004, ApJ 617, L9       &       79 	 & No        &	$<$1.2         & $-21.0<M_B<-19.0$	 &  general    \\
De Propris et al. 2005, AJ 130, 1516&      176   	 & No        &	$<$0.25        & $-21.7<M_B<-18.7$	 &  general    \\
Bell et al. 2006, ApJ 640, 241     &        6 	 & Yes       & 	0.1--0.7       &      $M_V<-20.5$	 &  dry        \\
De Propris et al. 2007, ApJ 666, 212&      112 	 & Yes	     &  0.01--0.12     & $-21.7<M_B<-18.7$	 &  general    \\
Kartaltepe et al. 2007, ApJS 172, 320&     1749 	 & No 	     &  0.1--1.2       &       $M_V<-19.8$	 &  general    \\
Lin et al. 2008, ApJ 681, 232      &      506 	 & No 	     &  0.1--1.2       &  $-21.0<M_B<-19.0$	 &  wet/dry    \\
Wen et al. 2009, ApJ 692, 511      &     1209 	 & Yes 	     &  $<$0.12        &       $M_r<-21.5$	 &  dry        \\
\hline
\end{tabular}
}
\end{center}
\end{table}

Galaxy merging plays an important role in many processes of astrophysics,
such as growth of massive galaxies, active galactic nucleus activity,
formation of supermassive blackhole binary (SMBH) and gravitational wave
(GW) radiation. Merger rate is one of key quantities for these studies.
Previous studies show that the pair fraction varies in a range of 1\%--10\%
in the redshift range of $z= 0.2$--1.2. These merger rates were usually
calculated from the projected close pairs, very few previous authors have
carefully checked the merging fraction of a large sample of pairs (see
Table~\ref{tab1}). Hence, the large uncertainty results from either the
small samples or the contamination of unphysical pairs. The merger rate
should be determined using the fraction of physically merging galaxies,
rather than the fraction of galaxies in the projected close pairs.

To get accurate merger rate in the local universe, we constructed a complete
pair sample of luminous early-type galaxies of $z<0.12$ using the SDSS data
with the following criteria: 1) $13.5<r<17.5$; 2) $(u-r)>2.2$ and
$(g-r)>0.7$; 3) $M_r<-21.5$; 4) $7<r_p<50$~kpc. 1209 pairs were selected
from 87,889 luminous early-type galaxies.  We then extracted the SDSS
$r$-band images for all selected pairs and performed sky background
subtraction, and applied the GALFIT package (Peng et al. 2002, AJ 124, 266)
to account for the smooth symmetric contributions of the luminous in the
images. Quantitative identification for interaction signatures from the
residual images suggests that 249 (21\%) pairs are merging. Considering the
total number of luminous early-type galaxies, we found that about 0.8\% of
the galaxies are merging. We adopted an average merging timescale of
0.3$^{+0.2}_{-0.1}$ Gyr (Bell et al. 2006, ApJ 640, 241) and found the
comoving volume merger rate $R_{\rm g}=(1.0\pm0.4)\times10^{-5}~{\rm
Mpc^{-3}~Gyr^{-1}}$.

SMBH binaries can be formed in galaxy mergers. From identified mergers, we
got the chirp mass distribution of SMBH binaries following $\log[\Phi(\log M
/M_{\odot})]=(21.7\pm4.2)-(3.0\pm0.5)\log M/M_{\odot}$. A large number of
SMBH coalescences generate a stochastic GW background at frequency
$10^{-9}$--$10^{-7}$~Hz, which is a promising GW source by the pulsar timing
array. Following Jaffe \& Backer (2003, ApJ 583, 616), we found that the
strain spectrum of GW background is $h_c(f)\sim 10^{-15}(f/{\rm
yr}^{-1})^{-2/3}$, one magnitude greater than previous estimates. See Wen et
al. (2009, ApJ 692, 511) for details.

\end{document}